\documentclass[runningheads]{svmult}

\usepackage{makeidx}   
\usepackage{graphicx}  
\usepackage{subeqnar}  
\usepackage{multicol}  
\usepackage{physprbb}  
\makeindex             

\begin{document}
\title*{Detection of CDM Substructure}
\toctitle{Detection of CDM Substructure}
%
%
\titlerunning{Detection of CDM Substructure}
%
\author{C.S. Kochanek\inst{1}
\and N. Dalal\inst{2}}
\authorrunning{Kochanek \& Dalal}
%
%
\institute{Harvard-Smithsonian Center for Astrophysics, Cambridge, MA 02420
\and Dept. of Physics, UCSD, La Jolla, CA 92093 }

\maketitle              

\def\gtorder{\mathrel{\raise.3ex\hbox{$>$}\mkern-14mu
             \lower0.6ex\hbox{$\sim$}}}
\def\ltorder{\mathrel{\raise.3ex\hbox{$<$}\mkern-14mu
             \lower0.6ex\hbox{$\sim$}}}
\def\farcs{\hbox{$.\!\!^{\prime\prime}$}}

\begin{abstract}
The properties of multiple image gravitational lenses  require 
a fractional surface mass density in satellites of $f_{sat}=0.02$
($0.006 \ltorder f_{sat} \ltorder 0.07$ at 90\% confidence)
that is consistent with the expectations for CDM.  The characteristic
satellite mass scale, $10^6$-$10^9M_\odot$, is also consistent
with the expectations for CDM.
The agreement between the observed and expected density of CDM substructure
shows that most low mass galactic satellites fail to form stars, and this
absence of star formation explains the discrepancy between the number of
observed Galactic satellites and CDM predictions rather than any modification
to the CDM theory such as self-interacting dark matter or a warm dark 
matter component.
\end{abstract}

\section{Introduction}
The existence of a ``CDM crisis'' (e.g. Moore 2001) rests on three pillars.  First,
the central rotation curves of some dwarf and low surface brightness galaxies may
be inconsistent with the expectation of a central dark matter density cusp, as is
discussed by Swaters and McGaugh in these proceedings.  Second, the observed and
predicted magnitudes and distributions of the baryonic angular momentum may be
inconsistent, as discussed by Dekel and Burkert in these proceedings.  Third, the
Galaxy has far fewer satellites than expected for a CDM halo of its mass
(e.g. Moore et al. 1999, Klypin et al. 1999).  It is this third pillar of the
crisis which we will undermine with our present analysis.

\begin{figure}[th]
\begin{center}
\centerline{
  \includegraphics[width=.45\textwidth]{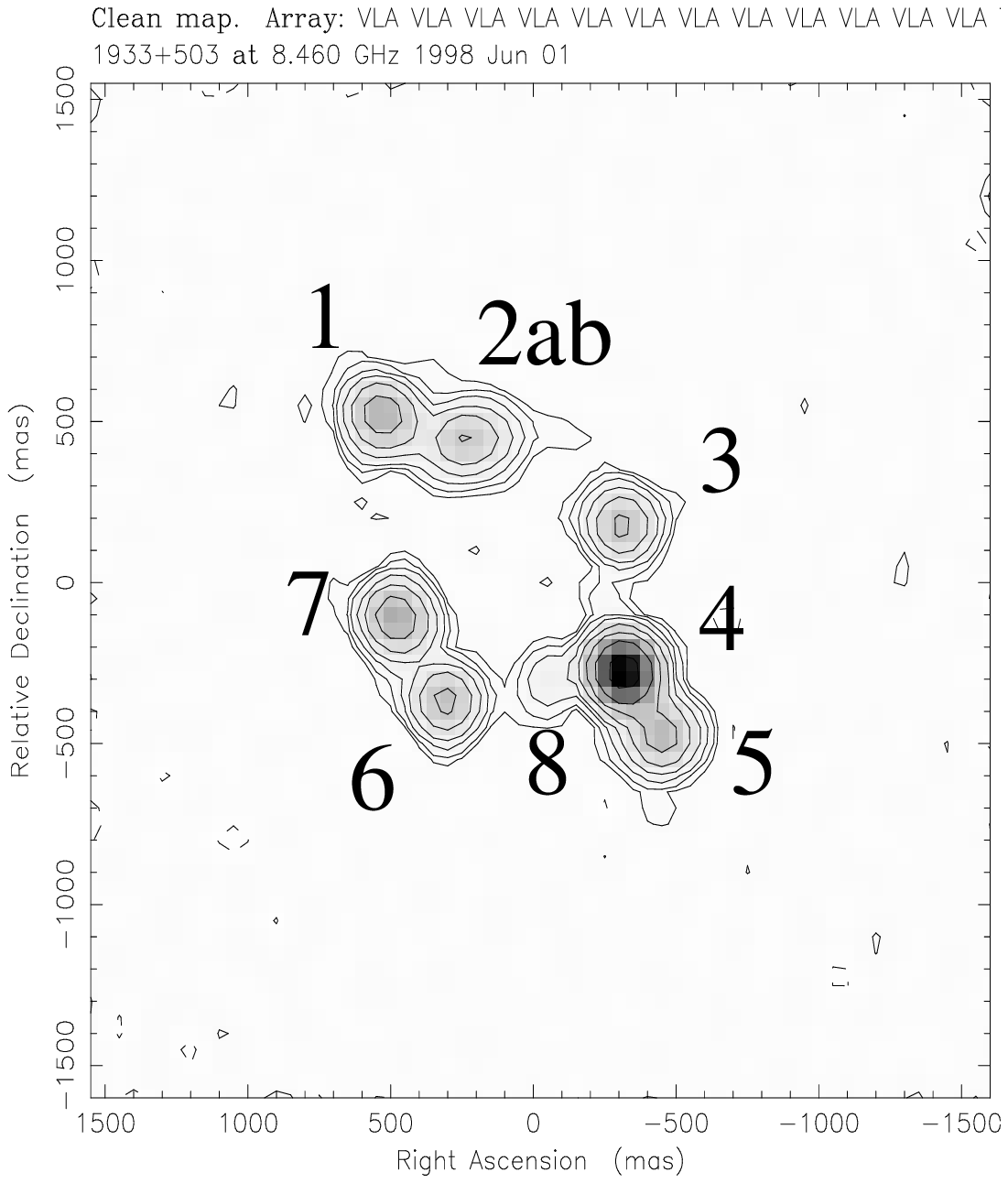}
  \includegraphics[width=.55\textwidth]{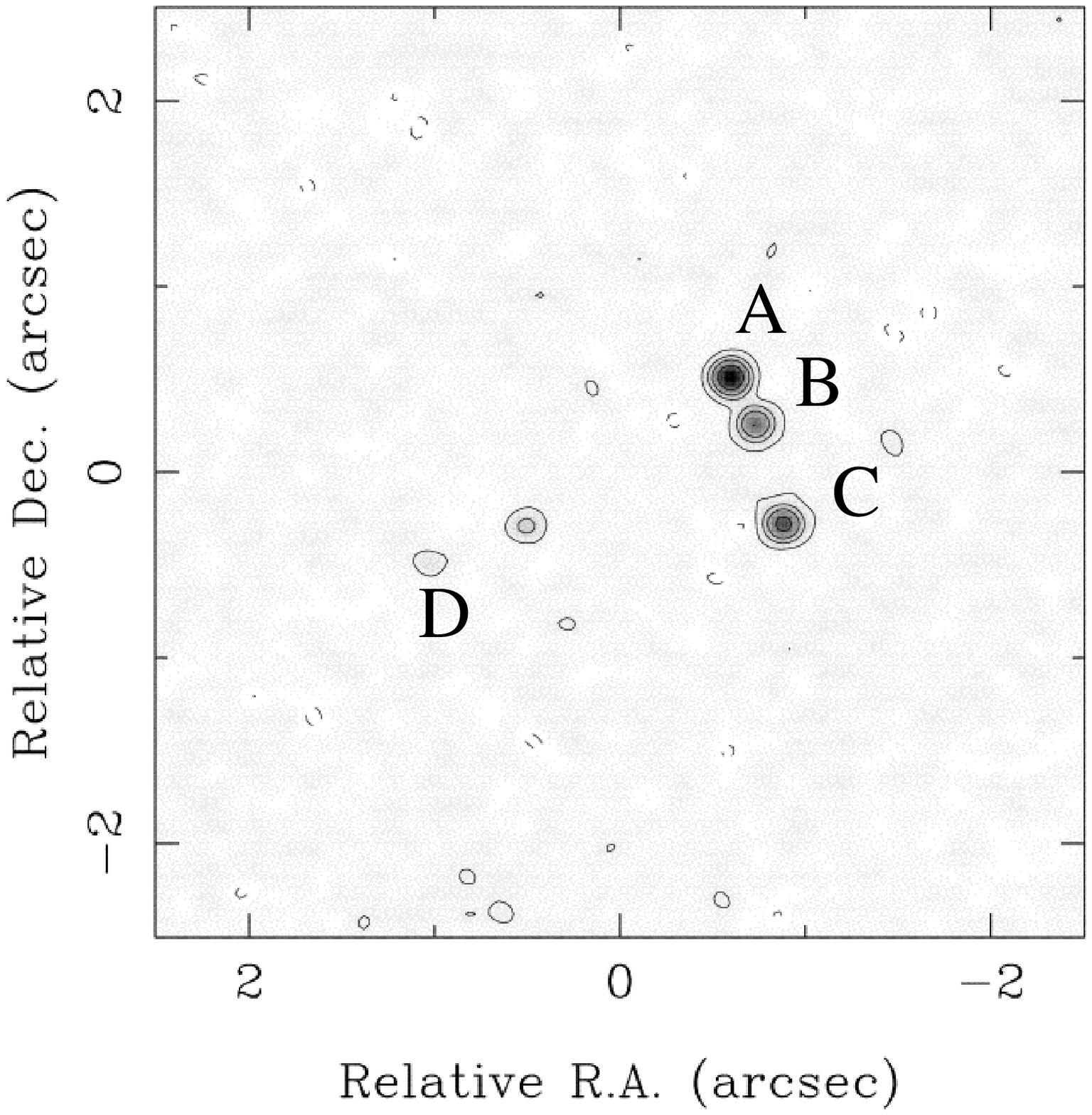}
  }
\end{center}
\caption[]{Examples of lenses with anomalous flux ratios:  
   the 4/136 images of B1933+503 (left, Sykes et al. 1998) and
   the A/B images of B2045+265 (right, Fassnacht et al. 1999).
   } 
\label{eps1}
\end{figure}

The satellite crisis essentially boils down to the observation that cluster and 
galaxy mass halos show similar amounts of substructure in CDM simulations, while observations
appear to show that the Galaxy and the Coma cluster have very different amounts of 
substructure.  The simplest solution is to suppress star formation in low mass
halos relative to high mass halos (e.g. Bullock et al. 2000).  A similar effect is
required to explain the difference between the steep slope of the
halo mass function ($dn/dM \sim M^{-2}$) and the shallow slope of the luminosity function
($dn/dL \sim L^{-1} \sim M^{-1}$) independent of the satellite problem (e.g.
Scoccimarro et al. 2001, Kochanek 2001, Chiu et al. 2001).  More exotic solutions
are to destroy the satellites using self-interacting dark matter (e.g. Spergel
\& Steinhardt 2000) or to avoid creating them by adding warm dark matter (e.g.
Bode et al. 2001, Colin et al. 2000).  We can distinguish between these 
possibilities only if we have a probe which is sensitive to the presence of
mass in the absence of light.

Moore et al. (1999) also realized that the only such probes we possess 
are gravitational lenses.  Indeed, Mao \& Schneider (1998) had already pointed
out that the anomalous flux ratios of close image pairs, where the fluxes 
expected for a mass distribution consisting only of the primary lens galaxy
would be nearly equal, could be explained by the presence of small, satellite 
galaxies.  More recently, Metcalf \& Madau (2001) showed that these effects would be easily
detected given the expected amount of substructure in CDM models, and 
Chiba (2001) showed that CDM substructure could explain the flux ratios
in PG1115+080 and B1444+231.  In Fig.~1 we show two less well-known
examples of gravitational lenses with anomalous flux ratios.
The problem, however, is to convert these arguments about plausibility
into quantitative estimates of the surface density and properties of 
satellites.

In Dalal \& Kochanek (2001) we developed a formalism for making 
these estimates and applied it to a sample of 7 four-image lenses.  
We do not repeat the mathematical development here, providing only
a qualitative outline in \S2.  We present our results and 
and discuss the future of the method in \S3.

\section{An Outline Of the Method}

The fundamental problem in estimating the properties of any substructure in
a gravitational lens is degeneracy between the effects of substructure and 
the primary lens.  For example, equal radial deflection perturbations to
the images are degenerate with a change in the mass of the macro model. 
Thus we must limit our analysis to systems with more constraints than
reasonable models of the primary lens have parameters.  This rules out
two-image lenses, so we will analyze only lenses consisting of four
images which have (in general) three degrees of freedom left after 
fitting a macro model consisting of a singular isothermal ellipsoid (SIE) in
an external shear field.

For each of the 7 lenses in our sample we start from the best fit 
model for the lens and linearly expand the lens equations 
around each image including both changes in the lens
parameters and the local perturbations due to substructure.
For any model of the substructure we can then adjust the parameters
of the macro model to compensate for its effects and compute a new
goodness of fit $\chi^2$.  This provides us with an estimate of the
probability $P(D_j|\delta_{ij})$ that substructure realization $i$
for lens $j$ provides a fit to the data $D_j$.  Of 
course, with so few degrees of freedom left after fitting the macro
model, many substructure realizations produce significant improvements
in the fit statistic -- we have too few constraints to uniquely
determine the substructure near each lensed image.  

\begin{figure}[ph]
\begin{center}
\centerline{
  \includegraphics[width=.5\textwidth]{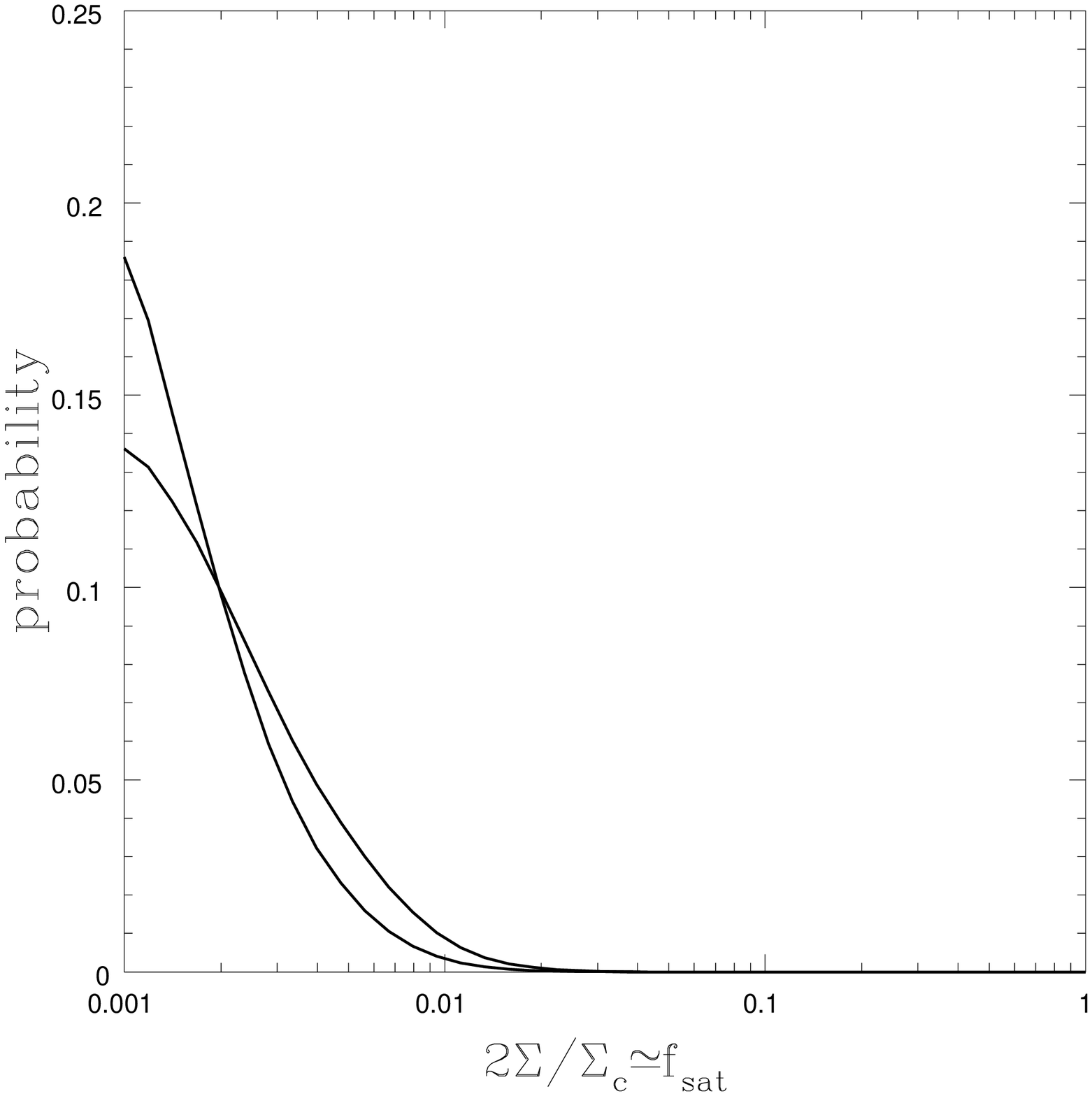}
  \includegraphics[width=.5\textwidth]{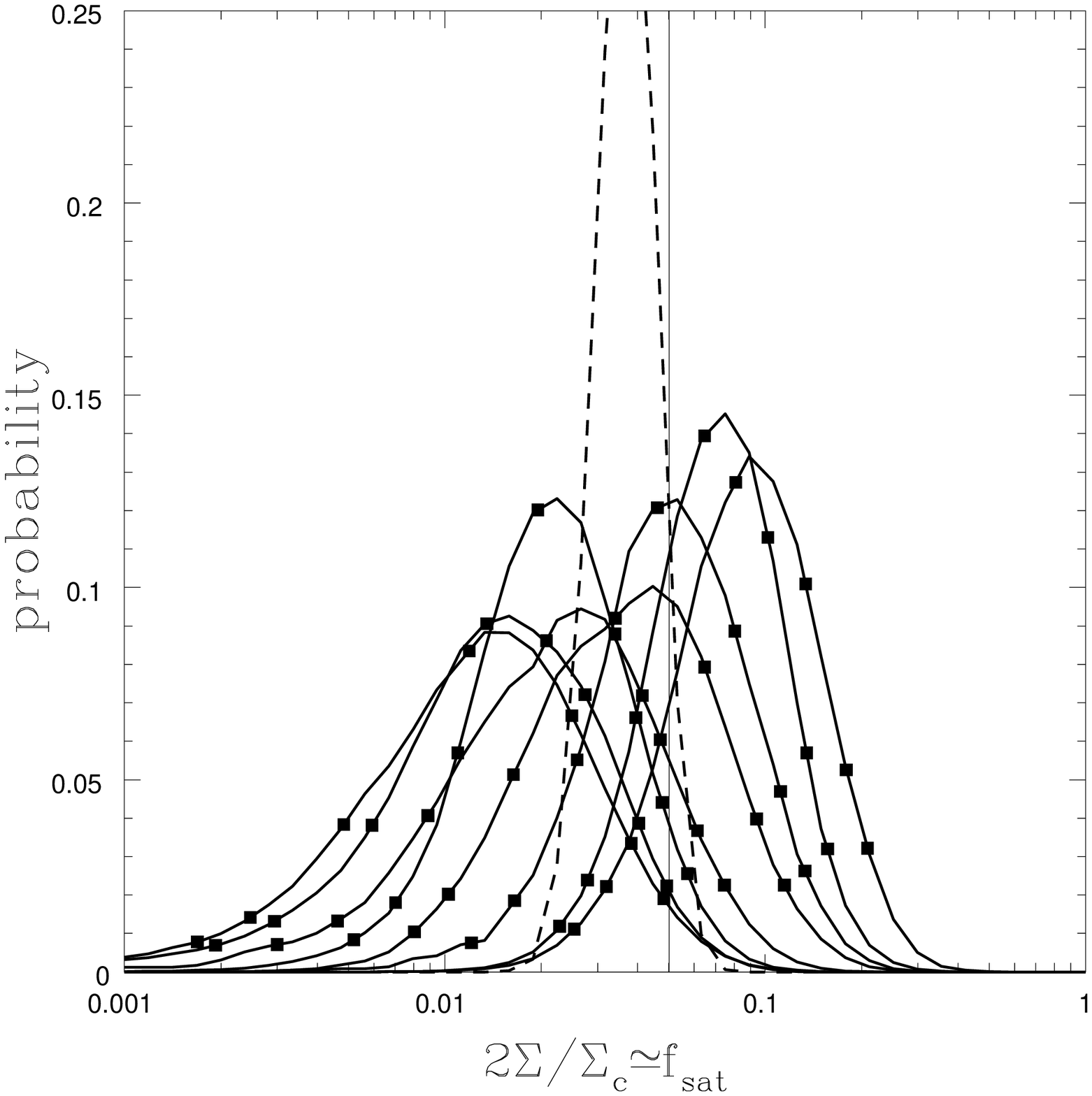}
  }
\end{center}
\caption[]{(left) Null tests.  The curves show the probability of
  fractional surface density $f_{sat}=2\Sigma/\Sigma_c$ for two
  Monte Carlo realizations of 7 lenses with measurement errors but
  no substructure.  }
\caption[]{(right) Monte Carlo realizations with $f_{sat}=0.05$
  and $b=0\farcs001$ modeled with $b=0\farcs001$.
  The solid curves show the probability for each of the 8 realizations, 
  and the dashed curve shows the joint probability for all 8 realizations
  (mimicking a sample of 56 lenses).  The points correspond
  to the median of the distribution and the range encompassing
  68\% (1$\sigma$), 90\% and 95\% (2$\sigma$) of the probability.
  The vertical line marks the true surface density.
  }
\begin{center}
\centerline{
  \includegraphics[width=.5\textwidth]{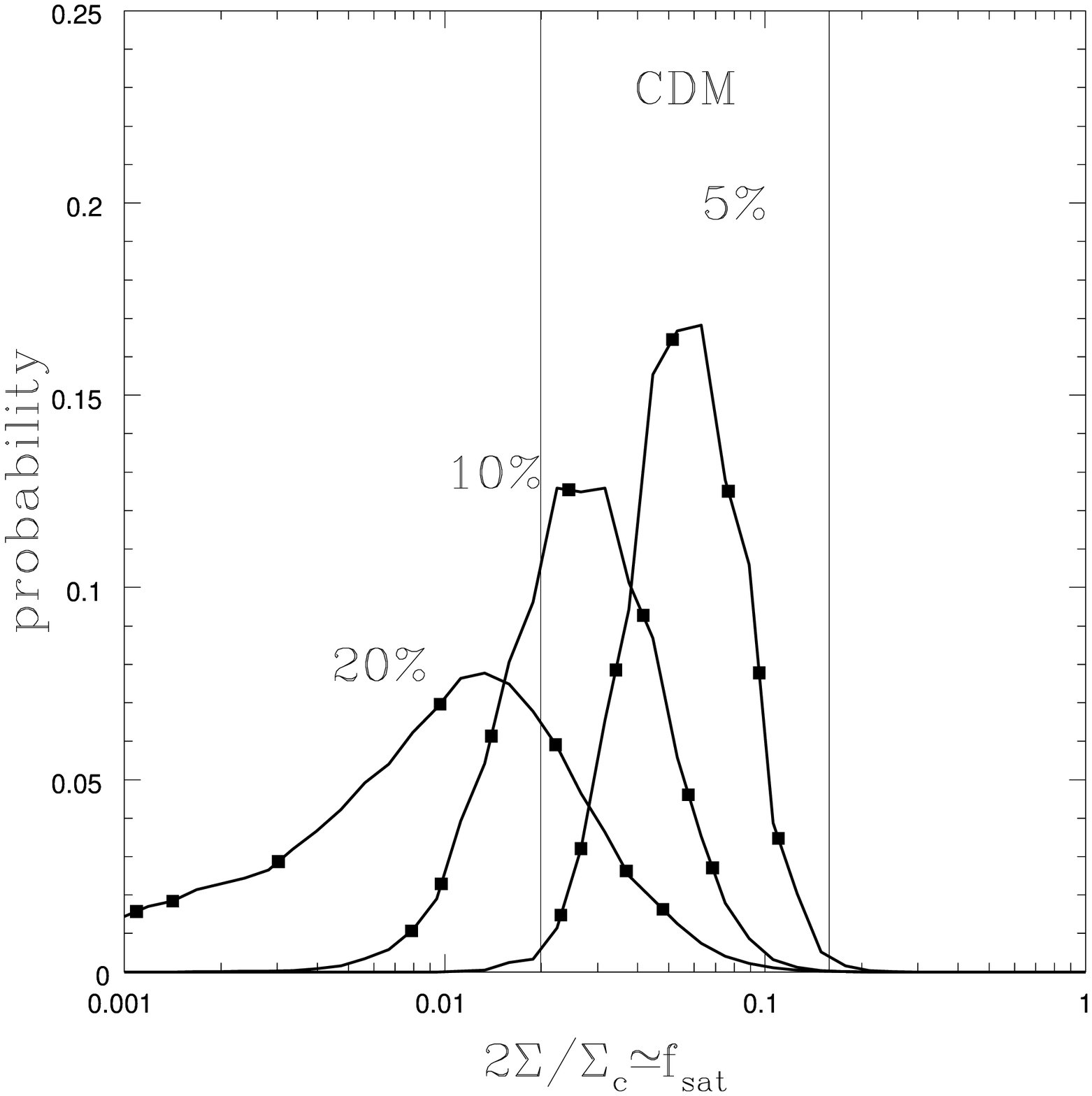}
  \includegraphics[width=.5\textwidth]{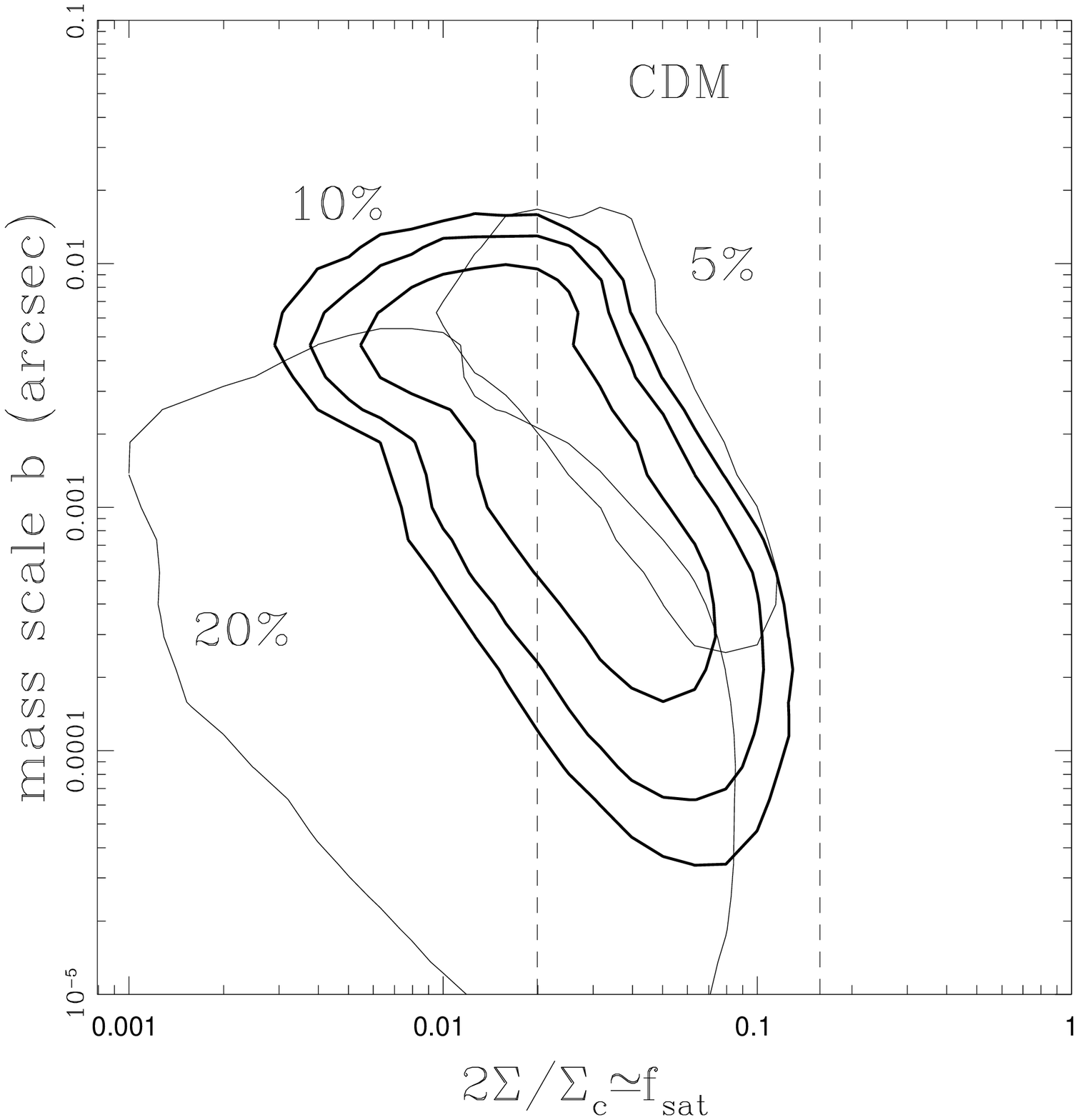}
  }
\end{center}
\caption[]{(left) Estimates of $f_{sat}$ in the real data for a fixed
  $b=0\farcs001$ mass scale assuming 5\%, 10\% or 20\% errors in the
  image fluxes. The points on the heavy curves correspond
  to the median of the distribution and the range encompassing
  68\% (1$\sigma$), 90\% and 95\% (2$\sigma$) of the probability.
  }
\label{data1}
\caption[]{(right) Simultaneous estimates of $f_{sat}$ and $b$ for the real
  data. For the 10\% flux error case (heavy lines) we show the 
  isoprobability contours encompassing 68\% (1$\sigma$), 90\% and 95\% (2$\sigma$)
  of the probability around the peak. For the 5\% and 20\%
  flux error cases, the light solid curve encompasses 90\% of the
  probability.
  }
\label{data2}
\end{figure}

\def\vp{\vec{p}}
We are interested in the statistical properties of the substructure, not
the particular realizations, and we can estimate the statistical properties
using a Bayesian analysis of the data.  For satellites drawn
from a population described by parameters $\vp$, the probability of 
substructure realization $\delta_{ij}$ is $P(\delta_{ij}|\vp)$.  Given
a prior $P(\vp)$ for the parameters, the probability of $\vp$
given the data after marginalizing over the actual realizations is
\begin{equation}
   P(\vp|D) \propto P(\vp)\Pi_j \sum_i P(D_j| \delta_{ij})P(\delta_{ij}|\vp)
\end{equation}
where the normalization is set by the constraint that 
$\int d\vp P(\vp|D)\equiv 1$ and we multiply the contributions from each
lens $j$ and sum (marginalize) over the substructure realizations $i$.
Qualitatively, as we vary the parameters, the probability of finding
substructure realizations which improve the fit from $\chi^2 \gg N_{dof}$
to $\chi^2 \sim N_{dof}$ varies, thereby allowing us to estimate the
parameters.

We modeled the substructure as pseudo-Jaffe models (Munoz et al. 2001) 
with surface densities, in units of the critical surface density $\Sigma_c$,
\begin{equation}
   \kappa = (b/2) \left[ r^{-1} - (r^2+a^2)^{-1/2} \right].
\end{equation}
The full model has three parameters: a mass scale $b$, a tidal truncation radius
$a$, and the satellite surface mass density $\Sigma$. For isothermal
models the tidal truncation radius is $a=(b b_0)^{1/2}$ where we fix
$b_0=1\farcs0$ for the Einstein radius of the primary lens.  The 
satellite mass is $M = \pi a b \Sigma_c$, and the fraction of the
mass in substructure is $f_{sat}\simeq 2\Sigma/\Sigma_c$.  Near the
Einstein ring, where we see the images, most of the mass is dark 
matter.  We model the substructure using random realizations of the
satellite distributions.  The perturbations are dominated by the
variance in the shear and convergence (rather than the astrometry),
with 
\begin{equation}
\langle \kappa^2 \rangle^{1/2} \simeq \langle \gamma^2 \rangle^{1/2}
   \simeq 0.13 \left( { 10 \Sigma \over \Sigma_c } \right)^{1/2}
        \left( { 10^3 b \over b_0 } \right)^{1/4}
        \left( { \ln \Lambda \over 10 }\right)^{1/2}.
\label{sheareq}
\end{equation}
There is a ``Coulomb'' logarithm $\ln \Lambda = \ln (a/s)$ where $s \ll a$ is
an effective core radius to the lens.

Figs.~2~and~3 show the results of two Monte Carlo tests of the algorithm.  In the
first test we take the macro models for our 7 lenses, add measurement errors and
analyze the data for substructure.  We find only upper bounds on the substructure
density with $f_{sat} \ltorder 0.004 $ and no preferred mass scale.  In the second
test we added perturbers with surface density $f_{sat}=0.05$,
mass scale $b=0\farcs001$, and tidal radius $a=0\farcs032$.  Multiple trials with
samples of 7 lenses recover the surface density and mass scale with reasonable
accuracy given the sample size.  In a third test, which we do not show, we find
that we can determine both $b$ and $f_{sat}$ simultaneously.

\section{The Surface Density of Satellites}

We then fit our sample of 7 lenses either with $b=0\farcs001$ (Fig.~\ref{data1})
or varying both $b$ and $f_{sat}$ (Fig.~\ref{data2}).  Qualitatively, the probability curves
look very similar to our Monte Carlo simulations.  The dominant uncertainty
is the degree to which the flux measurements in the real lenses are affected
by systematic errors.  To explore this, we assumed a standard flux uncertainty
of 10\%, but show results for both 5\% and 20\% flux errors.  The errors are
certainly smaller than 20\%, probably smaller than 10\% and unlikely to be
smaller than 5\%. In the models where we estimate both $f_{sat}$ and $b$
we find median estimates of $f_{sat}=0.020$ and $b=0\farcs0013$ for our standard 
model with 90\% confidence regions of $0.0058 < f_{sat} < 0.068$ and
$0\farcs0001 < b < 0\farcs007$.  If we assume 5\% errors, the surface
density must be higher, $0.013 < f_{sat} < 0.078$, while if we assume
20\% errors it must be lower, $0.0016 < f_{sat} < 0.051$.  There is a
relatively strong degeneracy between the values of $b$ and $f_{sat}$,
with higher mass scales allowing lower surface densities.  The slope
closely matches $b \propto f_{sat}^{-2}$, which corresponds to keeping
the rms shear perturbation constant (see eqn. \ref{sheareq}).

These estimates are consistent with the expectations for CDM models, where
$0.02 \ltorder f_{sat} \ltorder 0.15$ (Moore et al. 1999, Klypin et al. 1999),
and much larger than the expectations for normal satellite populations,
$10^{-4} \ltorder f_{sat} \ltorder 10^{-3}$ (Mao \& Schneider 1998, Chiba 2001).
For a $dn/dM \propto M^{-2}$ substructure mass function over the range
$M_{\rm low} < M < M_{\rm high}$, our mass scale $M=\pi a b \Sigma_c$ provides 
a crude estimate of the upper mass $M_{\rm high} \sim 10 M \sim 10^6M_\odot$ to
$10^9M_\odot$ which is also consistent with the CDM scenario.  Thus, our 
results are most naturally explained as a detection of the satellite
galaxies expected in the CDM model.

While one could debate whether we detect ``satellites'' or ``CDM satellites,''
alternative explanations do not appear to be viable.
There are systematic uncertainties in the data, but the anomalous flux ratios
which produce the result are generic and seen in repeated observations
spread over long time scales (years) and a range of wavelengths.  They
can be misinterpreted but not eliminated.  Coherent structures in the 
primary lens galaxy (e.g. spiral arms) are not seen in deep HST images
of the lenses, and would need to be far larger fractional mass perturbations
to produce the same effects because, unlike satellites, they cannot perturb 
individual images.  Stellar microlensing, while it is the same physical
phenomenon, is ruled out as a full explanation. Radio sources are 
generally too large to suffer from microlensing and the flux ratio
anomalies are too long lived.  While microlensing has been seen in one
radio lens (see Koopmans \& de Bruyn 2000), it only affects a superluminal
subcomponent, it has a small rms amplitude, and it has a very short
fluctuation time scale.  Moreover, our estimate of the
mass scale is consistent with that of satellites rather than stars.

There is considerable work to be done in the future.  First, there is no
reason that careful observation and monitoring of the lenses cannot measure
the image fluxes to 1\% accuracy including the effects of variability and
any microlensing.  Not only would this allow us to estimate the properties
of the CDM substructure more accurately, but it would also supply a large
sample of accurate time delay measurements for determining $H_0$ without
the problematic systematic errors of the local distance scale (e.g. 
Schechter 1999).  Second, deep high resolution imaging with both HST and
the VLBA is needed.  Images of the radio source's host galaxy can constrain
the macro model without being affected by substructure because of the 
larger angular size of the host.  Deep VLBA images to map extended 
structures can be used to measure the substructure mass scale accurately.
Extended radio structure also can be used to estimate the surface density
over larger areas of the lens than point sources, and can be used to
study the internal structure of the satellites.  Finally, more lenses
will reduce the statistical uncertainties.  This includes finding 
more lenses and more detailed studies of existing lenses to find
enough constraints on the macro model to allow a substructure analysis.

\noindent Acknowledgments: CSK was supported by the Smithsonian Institution
and NASA grants NAG5-8831 and NAG5-9265.  ND was supported by the Smithsonian
Institution Short Term Visitor Program, DOE grant DOE-FG03-97-ER 40546 and
the ARCS Foundation.

%

\end{document}